\documentclass[12pt,jpa,preprint]{iopart}
\usepackage{iopams}  
\usepackage{graphicx}

\begin{document}

\title[The role of geometry on dispersive forces]{The role of geometry on dispersive forces}

\author{C. E. Rom\'an-Vel\'azquez and Cecilia Noguez$^*$}

\address{Instituto de F\'isica, Universidad Nacional Aut\'onoma de M\'exico, Apartado Postal 20-364, M\'exico D. F. 01000, M\'exico}
\ead{$^*$cecilia@fisica.unam.mx}

\begin{abstract}
The role of geometry on dispersive forces is investigated by calculating the energy between different spheroidal particles and  planar surfaces, both with arbitrary dielectric properties. The energy is obtained in the non-retarded limit using a spectral representation formalism and calculating the interaction between the surface plasmons of the two macroscopic bodies. The energy is a power-law function of the separation of the bodies, where the exponent value depends on the geometrical parameters of the system, like the separation distance between bodies, and the aspect ratio among minor and major axes of the spheroid. 
 \end{abstract}

%\pacs{41.20.Cv, 77.55.+f, 02.70.Hm, 12.20.Ds}
%\maketitle

\section{Introduction}

The Casimir effect is one of the macroscopic manifestations of the fluctuations of the quantum vacuum~\cite{casimir}. Casimir showed that the energy $\mathcal{U}(z)$, between two parallel perfect conductor plates can be found from the change of the zero-point energy of the classical electromagnetic field, as
\begin{equation}
\mathcal{U}(z) = \frac{\hbar}{2} \sum_i [\omega_i(z) - \omega_i(z\to \infty)], \label{uint}
\end{equation}
where $\omega_i(z)$ are the proper modes that satisfy the boundary conditions of the electromagnetic field at the plates which are separated a distance $z$. The energy obtained by Casimir is a power law function of $z$, and at large distances $\mathcal{U}(z) \propto z^{-3}$, while at short distances $ \mathcal{U}(z) \propto z^{-2}$. Later, Lifshitz obtained a formula to calculate the force between two parallel half-spaces with arbitrary dielectric properties~\cite{lifshitz}. The Lifshitz formula depends only on the reflection amplitude coefficients of the half-spaces and the separation between them, finding the same dependence of the energy with $z$.
 In 1968, van Kampen \textit{et al.}~\cite{kampen} showed that the Lifshitz formula, in the non-retarded limit, is obtained from the zero-point energy resulting from the Coulomb interaction of the surface plasmons of the plates. After, Gerlach~\cite{gerlach} did an extension showing that also in the retarded limit, the Lifshitz formula is obtained from the interacting surface plasmons. Recently, it has been shown that the Casimir energy is given by the contribution from the interacting surface plasmons  and propagating modes in the cavity formed by the parallel plates~\cite{intravaia}. Furthermore, it has been shown that the contribution from the surface plasmons is essential to calculate the Casimir energy~\cite{intravaia}.

Surface plasmons are evanescent electromagnetic waves that propagate along the surface of conductors, and vanish elsewhere. By altering the surface, i.e., modifying size, shape, and/or environment of the conductor, the properties of surface plasmons can be tailored~\cite{noguezOP2005,gonzalez}. Therefore, if the shape of at least one of the bodies is modified, we would expect to observe changes on the energy due to the interaction between surface plasmons. In this paper, we study the influence of the geometry on the zero-point energy due to the interaction between macroscopic bodies. Using a method based on a spectral representation formalism \cite{ceci}, which determines the proper frequencies of the system, we calculate the zero-point energy  of all the interacting surface plasmons between a spheroidal particle and a flat plate, both with arbitrary dielectric functions. We find that the geometry plays an important role in the determination of the energy.

\section{Interacting surface plasmons}

We consider a spheroidal particle located near a flat substrate. The particle is generated by the rotation around one of the axes of an ellipse with lengths $2r_>$ and $2r_<$, with $r_> >r_<$. The symmetry axis of the particle is perpendicular to the substrate, and its center is located at a distance $d$ from the substrate which has a dielectric constant $\epsilon_{\rm sub}$. The particle has a dielectric function $\epsilon_{\rm part}$ and is embedded in an ambient of dielectric constant $\epsilon_{\rm amb} $, which is equal to 1 for vacuum. We consider that the three media: particle, substrate and ambient are non-magnetic. We are not considering nonlocal effects, therefore, all the variables are function of the frequency only. The explicit dependence of the following equations with the frequency is omitted here for simplicity. The charge distribution on the particle's surface, in the presence of a substrate, depends with the components of the external electromagnetic excitations, throughout the so called effective polarizability tensor $\bar{\alpha}_{\rm eff}$. When the particle is far from the substrate the effective polarizability becomes the polarizability of the isolated particle. But, when the particle is close to the substrate the induced multipolar interactions  modify the electromagnetic response of the system because more surface plasmon resonances of the particle interact with those in the substrate~\cite{noguezPRB04}. 

The quantum vacuum fluctuations induce a charge distribution on the particle which also induces a charge distribution in the substrate.  In the non-retarded limit,  the induced $lm$-th multipolar moment on the particle is given by~\cite{ceci} 
\begin{equation}
Q_{lm}(\omega)  =  \alpha_{\rm eff}^{lm} (\omega)\left[ V_{lm}^{\rm vac}(\omega) + V_{lm}^{\rm sub}(\omega) \right], \label{qlm}
\end{equation}
where $V_{lm}^{\rm vac}(\omega)$ is the field associated to the quantum vacuum fluctuations at the zero-point energy, $V_{lm}^{\rm sub}(\omega)$ is the induced field due to the presence of the half-space, and $\alpha_{\rm eff}^{lm}(\omega)$ is the $lm$-th component of the effective polarizability of the particle. 
It is known that the poles of $\bar{\alpha}_{\rm eff}$ yield the frequencies of the proper modes of the system~\cite{fuchs,bergman,milton}. In the non-retarded limit, the spheroidal radius of the major axis, and the minimun separation between the particle and the substrate are much smaller than the characteristic length of the system given, in this case, by $l=c/\omega_p$, with $c$ the speed of light, and $\omega_p$  the plasma frequency of the metallic particle. 

The analysis of $\bar{\alpha}_{\rm eff}$  for the system described above was done as follows. First, the electric potential induced in the system at any point in space was calculated to all multipolar orders. To find the solution for the induced potential a spectral representation of the Fuchs-Bergman-Milton type~\cite{fuchs,bergman,milton} was developed~\cite{roman}. By identifying the multipolar moments $Q_{lm}(\omega)$ induced in the particle, the components of $\bar{\alpha}_{\rm eff}$ were obtained. Then, the  frequencies of the proper modes for different shapes and locations of the particles is calculated by choosing a model for the dielectric function of the particle. For a detailed description of the method see Ref.~\cite{noguezPRB04}.
 
Within the spectral representation formalism, we can write  the component of $\bar{\alpha}_{\rm eff}$  in the following form: 
\begin{equation}
{\alpha}_{\rm eff}^{lm} (\omega, z) = - \frac{v}{4\pi} \sum_{s,q} \frac{C^{lm}_{sq}(z)}{u(\omega) - n^{lm}_{sq}(z)}, \label{alfa}
\end{equation}
where $v$ is the volume of the particle, $z$ is the minimum separation distance from  the particle to the substrate; such that, $z = d - r_>$ for prolate spheroids while $z = d - r_<$ for oblate spheroids. Here,
\begin{equation}
u(\omega)=[1-\epsilon_{\rm part}/\epsilon_{\rm amb}]^{-1}, \label{u}
\end{equation}
 is the so-called spectral variable and the strengths $C^{lm}_{sq}= (U^{lm}_{sq})^2$ are the so-called spectral functions where $U^{lm}_{sq}$ is the unitary matrix that satisfies the relation
\begin{equation}
 (U_{sq}^{l'm'})^{-1}\,H_{l'm'}^{s'q'}(z)\,U_{s'q'}^{lm}  =  4 \pi n_{sq}^{lm}(z) .
\end{equation} 
The matrix $H_{l'm'}^{s'q'}(z)$ depends only on the geometrical properties of the model and on the dielectric properties of substrate and ambient, through the contrast parameter 
\begin{equation}
f_c = (\epsilon_{\rm amb} -\epsilon_{\rm sub})/(\epsilon_{\rm amb} +\epsilon_{\rm sub} ). \label{fc}
\end{equation}
 Furthermore, $H_{l'm'}^{s'q'}(z)$   is a real and symmetric matrix which is given by~\cite{ceci}:
\begin{equation}
H_{lm}^{sq}(z) = n_{sq}^{lm}(z \to \infty) \delta_{sl} \delta_{qm} + f_c D_{sq}^{lm} (z),
\end{equation}
here $n_{sq}^{lm}(z \to\infty)$ are the depolarization factors of an isolated spheroid, and $D_{sq}^{lm} (z)$  is a matrix given by the multipolar coupling due to the presence of the substrate; this later vanishes when $z \to \infty$. Note that $H_{l'm'}^{s'q'} (z)$  contains all the information of the geometry of the system and the dielectric constant of the substrate, and is independent of the dielectric properties of the particle. In Eq. (\ref{alfa}), $\bar{\alpha}_{\rm eff}(\omega,z)$ is given as the sum of terms which show resonances at frequencies $\omega$ given by the poles of the equation, i.e., when $u(\omega) = n_{sq}^{lm}(z)$. As a consequence, an explicit procedure to calculate the strength and position of the resonances was obtained. In the next section we present results for the energy between a spheroid and a plate due to the interaction of all the induced multipolar surface plasmons.

\section{Zero-point energy of a spheroid near a substrate}

As follows, we present a systematic study of the energy for oblate and prolate particles with different asymmetries and considering different kinds of substrates. The energy is calculated by substituting the frequencies of the proper modes, obtained as described above, in Eq.~(\ref{uint}). Using the spectral representation, a systematic study of the energy in term of the geometrical parameters: $r_>$, $r_<$, and $z$, and the dielectric properties of the substrate, can be done. 

In the presence of a dielectric half-space, the proper modes, given by $\bar{\alpha}_{\rm eff}(\omega,z)$, are red-shifted always as the particle approaches the substrate, and this shift depends on the separation distance $z$. In general, the interaction energy is negative for any $z$, and is proportional to $(1+z)^{-\beta}$, with $\beta = 2L + 1$. Here $L $ is a positive integer which labels the highest order of the multipolar interaction, and depends on the geometrical parameters $r_>$, $r_<$, and $z$~\cite{noguezPRB04}.  For example, when $r_> = r_< =a$, and $z > 5 a$, then the relevant multipolar excitations are given by $L=1$, i.e., only surface plasmons with dipolar distributions are important. On the other hand, if $ 5a > z > 2a$ the interactions among surface plasmons with dipolar and quadrupolar charge distributions become relevant and $L =2$, such that, the energy, when $z\approx 2a$, is proportional to $(1 +z)^{-4}$. As $z \to 0$, more and more multipolar charge distributions must be taken into account, and when the spheroid is touching the substrate,  one would expect that $L \to \infty$, so that also the interaction energy also does.

As a case study, we employ the plasma model for the dielectric function of the particle,
$
\epsilon_{\rm part} (\omega )=1- \frac{\omega_{\rm p}^2}{\omega^2 }$,
where $\omega_{\rm p}$ is the plasma frequency, which is different for different metals. Therefore, the frequencies of the proper modes, obtained from the relation $u(\omega) = n_{sq}^{lm} (z)$, are 
\begin{equation}
\omega^{lm}_{sq}(z) = \omega_{\rm p} \sqrt{n^{lm}_{sq} (z)}, \label{modos2}
\end{equation}
and  according to Eq.~(\ref{uint}), the zero-point energy is given by
\begin{equation}
{\mathcal U}(z) = \frac{\hbar \omega_{\rm p}}{2} \sum_{sq,lm}\left[ \sqrt{n^{lm}_{sq}(z)} - \sqrt{n_{sq}^{lm} (z \to\infty) } \right], \label{ene2}
\end{equation}
where $n_{sq}^{lm}(z \to\infty)$ denotes the proper modes of the isolated spheroid. From eq.~\ref{ene2}, we can define a dimensionless energy $\Xi = {\mathcal U}(z)/\hbar \omega_p $, and study in detail the behavior of the system independently of the plasma frequency of the metallic spheroid.

\begin{figure}[htbp]
  \centerline{
  \includegraphics[width=\textwidth]{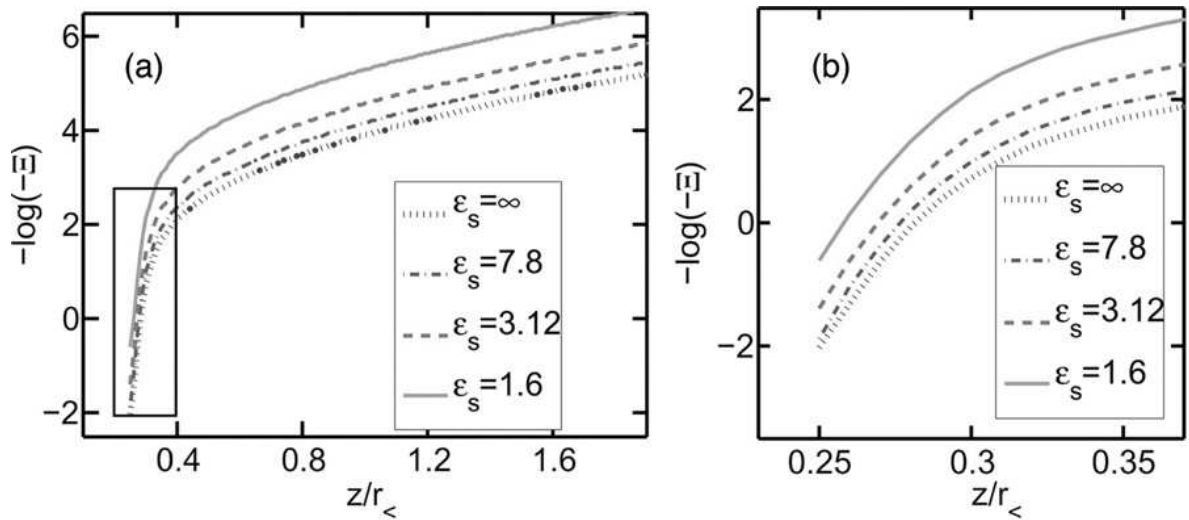} }
   \caption{Dimensionless energy as a function of $z/r_<$ for an oblate spheroid with $r_> / r_< = 1.4$ and different substrates.The inset in (a) is shown in (b).}
   \label{fig1}
\end{figure}
Let us first analyze the influence of the dielectric properties of the substrate on the energy as a function of the dimensionless distance $z/r_<$. In Fig.~\ref{fig1}, we show $-\log(-\Xi)$ for an oblate spheroid with an aspect ratio among the semi-axes of the spheroid, $r_> / r_< = 1.4$, and different substrates. Here, $\epsilon_{\rm sub} = \infty$ corresponds to a perfect conductor, while $\epsilon_{\rm sub} = 7.8$ and $3.12$ correspond to TiO$_2$ and sapphire, respectively. The case of $\epsilon_{\rm sub} = 1.6$ is just for illustration. The interaction strength between the particle and the substrate is modulate by the contrast factor given by Eq.~\ref{fc}, which is a function of $\epsilon_{\rm sub}$. For distances, $z/r_< > 1$, the energy goes to $ 0$  faster, as smaller is the value of $\epsilon_{\rm sub}$, as shown in Fig.~\ref{fig1}(a). At small distances, $z/r_< < 0.4$, the interaction between the particle and substrate increases considerably, because the largest value of the multipolar interaction $L $ becomes larger, as shown in Fig.~\ref{fig1}(b). However,  the value of $L$ is independent of $\epsilon_{\rm sub}$, and the observed differences at a given distance are due to the fact that the strength of the multipolar surface plasmon interactions are modulated with the factor $f_c$, such that, for larger values of the substrate dielectric constant the force strength increases. In Fig.~\ref{fig1}, we observe that the energy at a given distance is larger for greater values of $\epsilon_{\rm sub}$. The same general behavior of the energy as a function of the dielectric function of the substrate is also found for prolate spheroids.

\begin{figure}[htbp]
  \rightline{
    \includegraphics[width=1.00\textwidth]{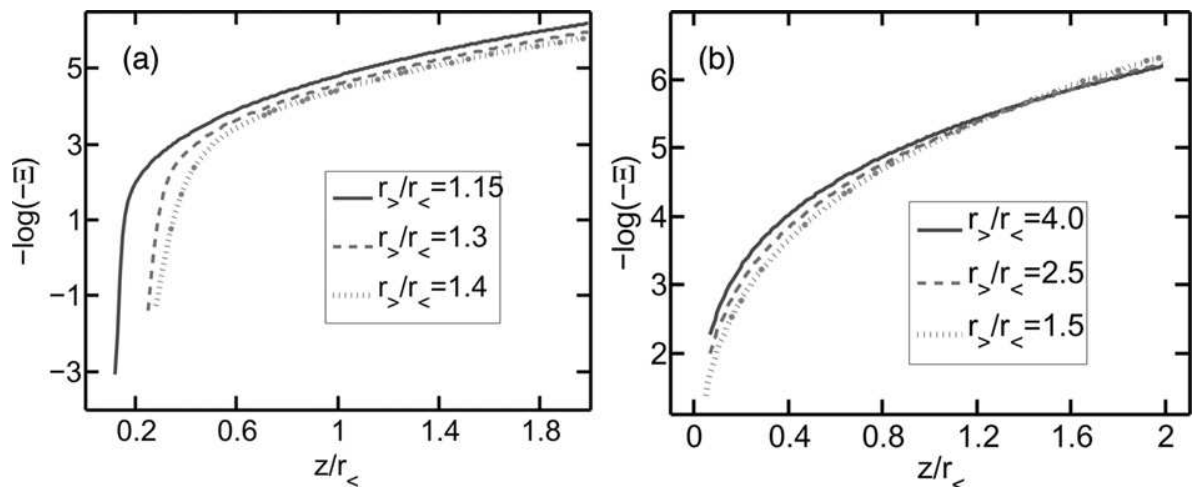} } 
       \caption{Dimensionless energy as a function of $z/r_<$ for  (a) oblate, and (b) prolate spheroids with different aspect ratios $r_> /r_<$.}
   \label{fig2}
\end{figure}

Now let us examine in detail the influence of the geometrical parameters on the energy between different spheroidal particles and a sapphire substrate, as a function of  the dimensionless distance $z /r_<$. In Fig.~\ref{fig2}(a), the dimensionless energy for oblate spheroids with different aspect ratios between the semi-axes, $r_> /r_< $, is shown. At a fix distance, the energy for oblate spheroids is larger  when the aspect ratio $r_> /r_<  \to 1$, i.e., when the spheroids tend to the spherical shape. Since the substrate is always the same, this means that the value of $L$ becomes larger when $r_> /r_< \to 1$ at any distance. In Fig.~\ref{fig2}(b), the same is shown for prolate spheroids, however,  we observe two different regimes in this case. When $z/r_< < 1.2$, the value of $L$ becomes larger when $r_> /r_< \to 1$, but the contrary occurs when $z/r_< > 1.2$. This means that the value of $L$ not only depends on the separation distance, but also on the axes aspect ratio  $r_> /r_<$, and the specific symmetry of the particle. From Fig.~\ref{fig2}, we also observe that the energy behavior is different for oblate and prolate spheroids, which is more evident when $z/r_< < 0.4$. The energy for oblate particles becomes more negative faster than for prolate ones, because the value of $L$ is larger for oblates. In conclusion, the exponent $\beta$ of the power-law behavior of the energy depends on the specific geometrical parameters  $z$,  $r_>$ and  $r_<$.

\begin{figure}[htbp]
  \rightline{
    \includegraphics[width=\textwidth]{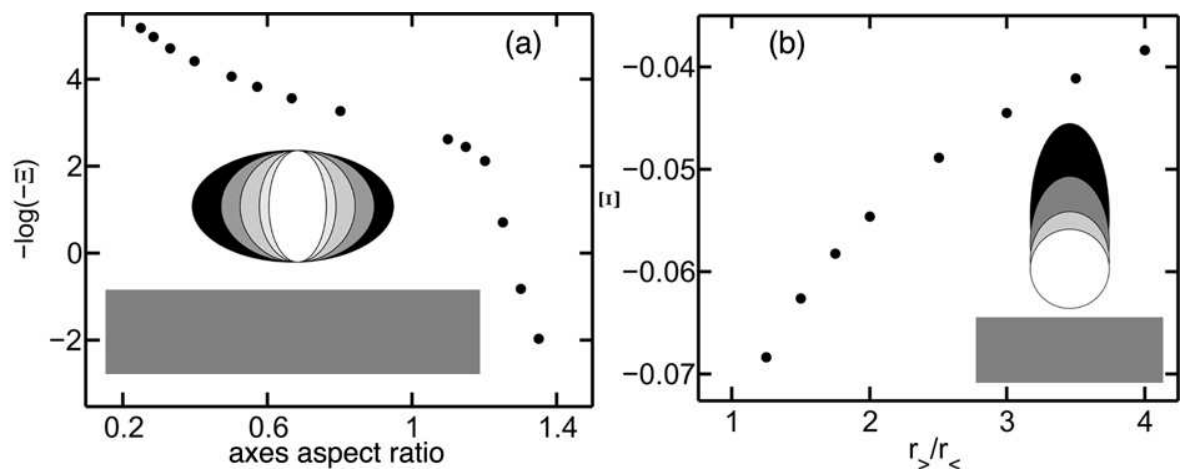} } % requires the graphicx package
   \caption{(a) Dimensionless energy of spheroidal particles at a fix distance $z$, as a function of the aspect ratio between the axis.  (b) Dimensionless energy of prolate spheroids at a fix $z$ and $r_<$ constant, as a function of $r_> /r_<$.}
   \label{fig3}
\end{figure}

To examine in detail the energy dependence with the geometry of the system, let consider different spheroidal particles at a fix distance $z$ and all with the axis normal the surface $r_\bot$, such that, $z/r_\bot = 0.25$ is a constant. In Fig.~\ref{fig3}(a) is shown the dimensionless energy as a function of the aspect ration $r=r_{||}/r_\bot$ between the axis parallel to the surface ($r_{||}$) and the one normal to it. The case when $r   < 1$ corresponds to prolate particles, while $r =1$ is the value for spheres, and  when $r   >1$  we have oblate ones, as it is illustrated in the inset of  Fig.~\ref{fig3}(a). Here, one can observe clearly the energy reliance on the geometry. For prolate spheroids the relevant multipolar interactions do not change dramatically with the aspect ratio of the axes. This means that the value of $L$ increases smoothly as the aspect ratio goes to $r=1$. On the other hand,  for oblate particles the number of multipolar interactions is more sensitive to $r$, and the values of the largest multipolar interaction rises dramatically with small increments of $r$.

In Fig.~\ref{fig3}(b),  the energy of prolate particles, at a fix $z$ and $r_<$ constant, with $z/r_> = 0.1$, as a function of $r_> /r_<$ is shown. In this case,  the projected area of the particle over the substrate is always the same, since the minor axis is constant for all cases (see inset in Fig.~\ref{fig3}(b)). Here, we observe that when the major axis increases the energy decreases because the reduction of number of relevant multipolar interactions involved, yielding different exponent of the power law function of $z$ for the energy. Therefore, we have shown that the energy is very sensitive to the geometry of the system.

\section{Discussion}

The above results might provide an insight into the range of validity of the Proximity Theorem Approximation, which was developed by Derjaguin and Abrikosova~\cite{proximidad} to estimate the force between two curved surfaces of radii $R_1$ and $R_2$. The Proximity Theorem Approximation assumes that the force on a small area of one curved surface is due to locally ``flat'' portions on the other curved surface. Within the Proximity Theorem Approximation  and when $z \ll R_1, R_2$, the force per unit area between two curved surfaces is:
\begin{equation}
\mathcal{F}(z) = 2 \pi \left(\frac{R_1R_2}{R_1 + R_2}\right) {\mathcal V}(z),
\end{equation}
 where $\mathcal{V}(z)$ is the Casimir energy per unit area between parallel plates separated by a distance $z$. In the limit, when $R_1 \to \infty$ and  $R_2 =R$, the problem reduces to a curved surface of radius $R$ and a flat plate, yielding 
\begin{equation}
\mathcal{F}(z) = 2 \pi R \mathcal{V}(z). \label{ptf}
\end{equation}
 The force obtained  is a power law function of $z$, and at ``large'' distances $\mathcal{F}(z) \propto z^{-3}$, while at short distances $ \mathcal{F}(z) \propto z^{-2}$. The force is proportional to the radius of the curved surface, and to the inverse of its projected area on the flat plate. Therefore, particles at a given distance with the same radii of curvature  would feel a force with the same power law function of $z$. Curved surfaces with the same projected area at a given distance would feel also a force with the same power law function of $z$

\begin{figure}[htbp]
  \rightline{
    \includegraphics[width=1.00\textwidth]{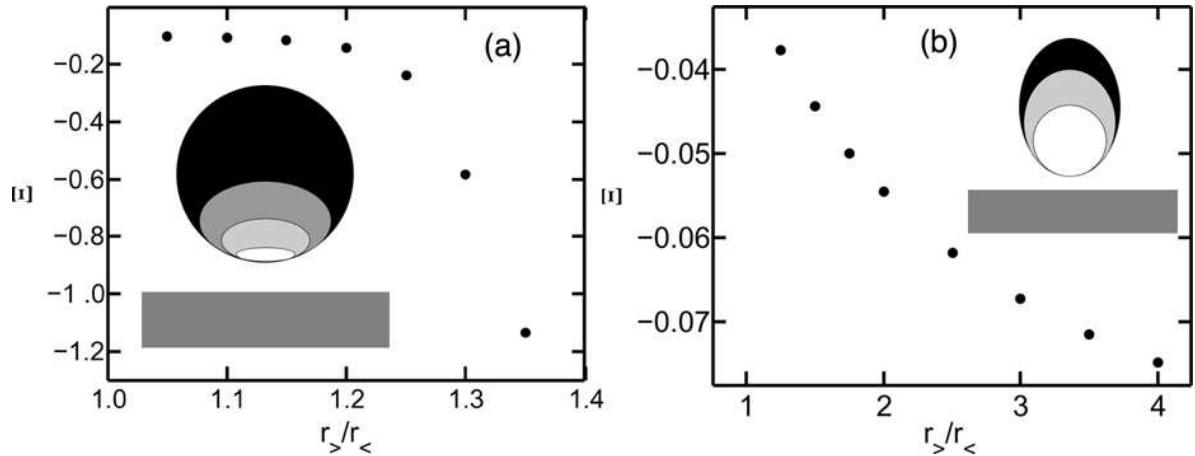} } % requires the graphicx package
   \caption{Dimensionless energy as a function of $r_> /r_<$ for  (a) oblate and (b) prolate spheroids with the same curvature, at a fix distance.}
   \label{fig4}
\end{figure}

 In Fig.~\ref{fig4}, we show the dimensionless energy as a function of the aspect ratio $r_> / r_<$, for (a) oblate and (b) prolate particles. We consider the case when all oblate (prolate) particles have the same curvature and are at a distance $z/r_> = 0.25$ ($0.1$). The energy for oblate, as well as, for prolate particles is a power law function of $z$, which depends on the geometrical parameters, even when the curvature of the particle is not changed. In the case of prolate particles, the value $L$ increases smoothly as the aspect ratio also does. On the other hand,  for oblate particles the number of multipolar interactions is more sensitive to axes aspect ratio, and the values of the largest multipolar interaction increases dramatically as $r_> / r_<$ also does. 

\section{Conclusions}

The role of geometry in dispersive forces in the non-retarded limit is studied by calculating the energy from the interacting surface plasmon of macroscopic bodies. We analyze in detail the interaction of oblate and prolate particles with a substrate. When the particle is close to the substrate the multipolar interactions induced by the substrate modify the electromagnetic response of the system. In general, we find that the energy is described by a power law function whose exponent depend on the minor and major axes of the spheroid, as well as, with the separation between bodies.  

\subsection*{Acknowledgments}
We acknowledge the partial financial support from CONACyT Grant No. 44306-F, and from DGAPA-UNAM Grant No.~IN101605

\end{document}